\documentclass[conference,10pt]{IEEEtran}
\usepackage[utf8]{inputenc}
\usepackage[T1]{fontenc}
\usepackage{mathtools}
\usepackage[thinc]{esdiff}
\usepackage{commath}
\usepackage{mathtools}
\usepackage{amsthm}
\usepackage{dsfont}
\allowdisplaybreaks
\usepackage{amsmath,amsfonts,amssymb}
\usepackage[ruled, linesnumbered]{algorithm2e}
\usepackage[noend]{algpseudocode}
\usepackage{float}
\usepackage{graphicx}
\usepackage{mathtools}
\usepackage[thinc]{esdiff}
\usepackage{epsf}
\usepackage{epsfig, amsmath, amssymb, graphicx}
\usepackage{subcaption}
\usepackage{multirow}
\usepackage{cite}
\usepackage[english]{babel}
\DeclareGraphicsExtensions{.pdf, .bmp, .eps, .ps, .jpeg, .png}
\long\def\comment#1{}

\usepackage{diagbox}

\title{Towards Practical Privacy-Preserving Solution for Outsourced Neural Network Inference} 
\author{Pinglan Liu, Wensheng Zhang\\
Department of Computer Science, Iowa State University, 
Ames, Iowa, USA 50011\\
E-mail: \{pinglan,wzhang\}@iastate.edu}

\begin{document}
\maketitle

% \begin{abstract}
% \end{abstract}

\begin{abstract}
When neural network model and data are outsourced to cloud server for inference,
it is desired to preserve the confidentiality of model and data
as the involved parties (i.e., cloud server, 
model providing client and data providing client) may not trust mutually. 
Solutions were proposed based on  
multi-party computation, trusted execution environment (TEE) and 
leveled or fully homomorphic encryption (LHE/FHE),
but their limitations hamper practical application. 
%
% Hence, 
We propose a new framework based on synergistic integration of LHE and TEE,
which enables collaboration among mutually-untrusted three parties, while
minimizing the involvement of (relatively) resource-constrained TEE and 
allowing the full utilization of the untrusted but more resource-rich part of server.
We also propose a generic and efficient LHE-based inference scheme 
as an important performance-determining component of the framework. 
We implemented/evaluated the proposed system on a moderate platform and 
% conducted extensive evaluations to 
show that, our proposed scheme is more applicable/scalable to various settings,
and has better performance,  
compared to the state-of-the-art LHE-based solutions.
% which are also more restrictive in applicability and scalability. 
\end{abstract}

% \vspace*{-0.1in}
% {\noindent\bf Keywords: } Outsourcing, Confidentiality, CNN, Homomorphic Encryption. 

% \vspace*{-0.1in}

\section{Introduction}
\label{sec:intro}

As deep neural network based machine learning often demands
large computational resource and data,
outsourcing computation and data to cloud servers is popular. 
% With the outsourcing, 
Outsourced data and neural network model can be exposed or stolen 
when they are at untrustable or vulnerable servers. 
Hence, we should protect their confidentiality; meanwhile, 
such protection should not prevent computations by authorized parties.

Extensive research has been conducted on protecting confidentiality of outsourced data and/or model.
The adopted approaches roughly fall into the categories of 
multi-party computation (MPC), 
trusted execution environment (TEE), and
leveled or fully homomorphic encryption (LHE/FHE). 
The MPC-based solutions~\cite{mohassel2017secureml, rouhani2018deepsecure, liu2017oblivious, juvekar2018gazelle, mishra2020delphi, chandran2019ezpc, riazi2019xonn, riazi2018chameleon,mohassel2018aby3, wagh2020falcon,baryalai2016towards}  
often also leverage partly homomorphic encryption.
They have better computational efficiency,
but at the expense of higher communication overhead and network latency, 
since multiple parties need to interact with each other 
while computation is being conducted. 
Some of them also have to assume no collusion between parties~\cite{patra2020blaze,koti2021swift,chaudhari2019trident,byali2019flash}. 
TEE-based solutions~\cite{zhang2021citadel,natarajan2021chex} 
employ the hardware-based technologies such as Intel SGX
to set up secure enclaves at the server to perform outsourced computation. 
% In spite of the high level of security attained, 
Resource (e.g., memory) available at TEE is (relatively) constrained\footnote{Taking Intel SGX as example,
common CPUs have up to 128MB SGX memory;
though some recent-generation Intel CPUs~\cite{intelXeonSGX} 
support 8GB-512GB SGX memory,
the size is still much smaller than that of their regular memory (up to 6TB); also, 
TEE has much higher paging overhead due to encryption.},
but the much more rich resources (including computationally-powerful GPUs) cannot be fully utilized.
Thus, the scalability of such solutions can be limited. 

LHE/FHE, which enables computation over encrypted data,
is a promising tool for protecting data and model confidentiality while 
preserving its utility even in untrusted environment.
Also, it does not require communication between multiple parties during computation and thus
avoids the overhead and complexity due to communication; 
it can make use of the available computational (including GPUs) and memory resources.  
Its notoriously-high computation cost has long hampered it from being applied in practical systems,
but theoretical and practical advancements~\cite{brakerski2014efficient, simd, cheon2017homomorphic} 
in the last decade are gradually changing the landscape. 
In particular, Smart and Vercauteren~\cite{simd} proposed the packing technique, which 
indicates that, by packing a large number of values into the {\em slots} of a ciphertext, 
large-scale SIMD operations can be performed on these packed values at one time
and hence the amortized cost can be significantly reduced. 
Applying LHE/FHE along with the packing technique, 
numerous research works~\cite{gilad2016cryptonets, jiang2018secure, bourse2018fast, dathathri2019chet, xie2019bayhenn, xie2021privacy} 
have been recently reported to protect data and model confidentiality 
for inference based on convolutional neural network (CNN). 
Among them, CryptoNets~\cite{gilad2016cryptonets} and E2DM~\cite{jiang2018secure} are the most representative. 
However, CryptoNets assumes the availability of a large number of inputs to fill all the slots in each ciphertext. 
Though it has a high level of amortized efficiency,
this may not be attainable in practice, 
where a client may not have a large set of inputs available at the same time.
E2DM proposes a more sophisticated packing method that can efficiently utilize the slots 
when the number of simultaneously-available inputs is smaller. 
However, 
% perhaps due to its focus on optimizing matrix multiplication,
it only considers the CNN model with one convolutional layer and 
no generic method is provided for generic CNN models.

To address the limitations with the state of the art,
we propose more practical framework and schemes to preserve 
the confidentiality of both neural network model and data
outsourced to an untrusted server for inference. 
The main contributions of our work are as follows.

First, we propose a new framework that applies both LHE and TEE. 
With the framework, 
a cloud server first sets up a TEE,
which initializes a LHE system.
After attesting the TEE, 
a model providing client (i.e., model provider) 
uploads to the server its model parameters 
encrypted with the LHE public key provided by the TEE, and 
a data providing client (i.e., data provider)
uploads to the server its input data 
encrypted with the same public key.
The untrusted part of the server (i.e., REE - the resource-rich execution environment out of TEE),
conducts the computation based on the encrypted model and data,
and sends the inference result to the TEE,
which decrypts the result, re-encrypts it with the data provider's secret key
and sends it to the data provider.
The proposed framework has the following advantages: 
it enables secure interactions among the three parties
(model provider, data provider and cloud server),
who may not trust mutually;
it minimizes the involvement of the TEE, which is (relatively) constrained in resource,
for only the most essential works of initializing the LHE system and 
decrypting/re-encrypting inference results;
it allows the REE to be fully utilized in performing the outsourced computation. 

Second, we propose a generic and efficient LHE-based inference scheme as 
an important performance-determining component of the proposed framework.
With the scheme, a CNN model with 
arbitrary numbers of convolutional and fully-connected layers can be supported. 
Also, the model provider can specify a desired size of simultaneously-available inputs
which can be much smaller than the total number of slots at a ciphertext.
New packing methods and propagation algorithms are devised to 
efficiently pack the model parameters and the simultaneously-available inputs into cipherhexts
and then efficiently process them, such that: 
ciphertexts are processed smoothly through all layers of a CNN model 
without decryption or re-encryption; 
the slots in ciphertexts are utilized as much as possible;
the encryption level is minimized to reduce 
the size and processing complexity of ciphertexts. 

Third, we implemented the proposed system and 
evaluated it extensively on a moderate platform. 
The evaluation results reported in this paper focuses on the following:
% \begin{itemize}
%     \item 
    Our system has good scalability with varying size of simultaneously-available inputs;
    the computational efficiency improves as the size increases, meanwhile
    it does not degrade significantly as the size decreases.
    % \item 
    Our system is applicable to various configurations of CNN model.
    % \item 
    Our system outperforms the state-of-the-art LHE-based schemes~\cite{gilad2016cryptonets, jiang2018secure}, 
    which is the most related to our work,
    in addition to that our system is applicable in more generic and practical settings.
% \end{itemize}

In the rest of the paper,
we present the background in Section II,
our proposed framework in Section III, and
our LHE-based inference scheme in Section IV.
% and optimizations in Section V.
Section V reports our evaluation results,
and Section VI briefly surveys the related works.
Finally, Section VII concludes the paper.

\section{Preliminaries}
\label{sec:prelim}

\subsection{System Model and Problem Description}
\label{subsec:syst-model}

We consider a system composed of a cloud server and multiple clients.
The server consists of 
trusted execution environments (TEEs)
each could be an Intel SGX~\cite{mckeen2013innovative} enclave that has been successfully attested,
and a untrusted but resource-rich execution environment (called REE, 
i.e., execution environment outside of any SGX enclave). 
Note that, side-channel attacks to TEE and the corresponding mitigation techniques have been extensively studied 
and are out of the scope of this paper. Deploying any mitigation technique 
is orthogonal to and thus can be integrated with our proposed design.  
Also, as our design tries to minimize the involvement of TEE (i.e.,
TEE is only used for key generation/distribution and the decryption/re-encryption 
of inference results), we anticipate that deploying mitigation techniques should not 
affect the system performance much.   

There are two types of clients, 
a model provider and multiple data providers.
The model provider owns a neural network model 
(e.g., a CNN model for image classification) that 
it has already trained.
Each data provider has data (e.g., images) that 
can be processed with the neural network model. 
The clients want the cloud server to use the model to make inference based on the data.
Meanwhile, data and model privacy should remain confidential. 
Specifically, the data should not be revealed to anyone 
other than its providing client
and the model parameters should not be revealed to anyone 
other than its providing client.
However, we allow the hyper-parameters of the model 
(including the number of layers and the number of nodes on each layer) 
to be revealed. 

% The client determines the hyper-parameters (e.g., the topology of the network) 
% of the CNN model, and these parameters are shared with the server. 
% It owns the inference data, and outsources the data (after encrypted with leveled FHE)
% to the server for model inference. 

% The leveled FHE is asymmetric encryption. 
% The private key is shared only by the client and the TEEs at the server,
% while the public key is published to all the parties 
% (including the REE at the server). 

\subsection{CNN Model}
\label{subsec:cnn-model}

For a CNN model outsourced by its provider,
we assume that it consists of  
a sequence of $c$ convolutional layers (called CLs) followed by 
$f$ fully-connected layers (called FLs).

For each convolutional layer $l\in\{0,\cdots,c-1\}$,
we assume that it has $\alpha_l$ channels,
one input matrix for a channel is of $\beta_l\times\beta_l$ elements
(thus $\beta_l$ is {\em side} of an input matrix of the layer),
it has $\epsilon_l$ filters for each channel,
the side of a filter is denoted as $\gamma_l$ 
and the stride is denoted as $\delta_l$.
%
% If layer $i$ is a convolutional layer,
% there are $\epsilon_i$ filters for each channel 
% and each of the filters has the side of $\gamma_i$.
%
% If layer $i$ is a pooling layer, 
% a certain non-linear function, 
% denoted as $p_i(x_0,\cdots,x_{\gamma_i^2-1})$
% is used to conduct the pooling operation over
% all the elements of each kernel.
%
For each fully-connected layer $l\in\{0,\cdots,f-1\}$,
the numbers of input and output neurons are denoted as $\iota_l$ and $o_l$, respectively.
Thus, its weight matrix, denoted as $M^{(l)}$, has dimensions $\iota_l\times o_l$.

For simplicity, we assume the square function is used as the activation function for 
each layer, as in \cite{jiang2018secure,gilad2016cryptonets}.
Note that, there have also been extensive studies on the evaluation of various activation functions over encrypted data~\cite{lu2017cryptii, chou2018faster, lu2021pegasus}.
Such schemes can be integrated with our proposed scheme, but do not elaborate on this as
it is out of the scope of this paper.

% For model training,
% suppose $n$ sets of training data are processed per training epoch. 
% %
% Each FHE ciphertext has $s$ slots to encode/encrypt up to $s$ plaintexts. 

\subsection{Homomorphic Encryption Primitives}

We use an asymmetric leveled homomorphic encryption (LHE) scheme, 
which enables the computation on encrypted data (i.e., ciphertexts) 
% without decryption 
% such that the encrypted result matches that of the computation on plaintext,
as long as the computations do not exceed a certain predefined {\em level}.
We also assume that the scheme adopts the ciphertext packing technique, 
with which multiple values can be encoded and encrypted into a single ciphertext
and operations 
% (including addition and multiplication between ciphertexts or between a plaintext and ciphertext)
can be performed in a SIMD manner. 
% Moreover, we assume that rotation operation can be performed on 
% the values that are encrypted into a ciphertext.  
Formally, the LHE scheme has the following primitives:

\begin{itemize}
    \item 
$(sk,pk,evk)\leftarrow KeyGen(1^\lambda, {\cal L})$: 
given security parameter $\lambda$ and the highest level of encryption ${\cal L}$, 
    it outputs a secret key $sk$, a public key $pk$, an evaluation key $evk$ and
    the the slot number ${\cal S}$ for each ciphertext. Here,
    ${\cal S}$ is also the number of scalar values that can be encoded and encrypted to a ciphertext.

    \item
$ct\leftarrow Enc_{pk}(\vec{pt})$: 
given public key $pk$ and 
    a plaintext vector $\vec{pt}=(pt_0,\cdots,pt_{{\cal S}-1})$, 
    it outputs ciphertext $ct$.

    \item
$\vec{pt}\leftarrow Dec_{sk}(ct)$: 
given secret key $sk$ and a ciphertext $ct$, 
    it outputs plaintext vector $\vec{pt}$ whose values are encoded and encrpyted to $ct$.

    \item 
$ct'\leftarrow ct_1\oplus ct_2$: 
given ciphertexts $ct_1$ and $ct_2$, it outputs ciphertext $ct'$ s.t. $Dec_{sk}(ct')=Dec_{sk}(ct_1)+Dec_{sk}(ct_2)$. Note that, the $+$ operator stands for element-wise 
    addition between two vectors; that is, 
    if $Dec_{sk}(ct_1)=\vec{pt}_1 = (pt_{1,0},\cdots, pt_{1,{\cal S}-1})$ and 
    $Dec_{sk}(ct_2)=\vec{pt}_2 = (pt_{2,0},\cdots, pt_{2,{\cal S}-1})$,
    then $Dec_{sk}(ct_1)+Dec_{sk}(ct_2) = (pt_{1,0}+pt_{2,0},\cdots, pt_{1,{\cal S}-1}+pt_{2,{\cal S}-1})$.

    \item 
$ct'\leftarrow ct_1\otimes ct_2$: 
given ciphertexts $ct_1$ and $ct_2$, it outputs ciphertext $ct'$ s.t. $Dec_{sk}(ct')=Dec_{sk}(ct_1)\times Dec_{sk}(ct_2)$. Note that, the $\times$ operator stands for element-wise 
    multiplication between two vectors; that is, 
    if $Dec_{sk}(ct_1)=\vec{pt}_1 = (pt_{1,0},\cdots, pt_{1,{\cal S}-1})$ and 
    $Dec_{sk}(ct_2)=\vec{pt}_2 = (pt_{2,0},\cdots, pt_{2,{\cal S}-1})$, then 
    $Dec_{sk}(ct_1)\times Dec_{sk}(ct_2) = (pt_{1,0}\times pt_{2,0},\cdots, pt_{1,{\cal S}-1}\times pt_{2,{\cal S}-1})$.
    
    \item 
$ct'\leftarrow CMult(ct,\vec{pt})$: 
given ciphertext $ct$ and plaintext $\vec{pt}$, 
    it outputs ciphertext $ct'$ s.t. $Dec_{sk}(ct')=\vec{pt}\times Dec_{sk}(ct)$.
    
    \item 
$ct'\leftarrow Rot(ct,m)$: 
given ciphertext $ct$ that encrypts $\vec{pt}=(pt_0,\cdots,pt_{{\cal S}-1})$ and integer $m<{\cal S}$, 
    it outputs $ct'$ which is ciphertext for $(pt_{m},\cdots,pt_{{\cal S}-1},pt_0,\cdots,pt_{m-1})$.

\end{itemize}
Note that, the operations involving ciphertext,
i.e., $\otimes$, $CMult$ and $Rot$, 
use some keys which are skipped here for brevity. 
In our design and implementation, 
we adopt CKKS~\cite{cheon2017homomorphic}, 
which provides all the above primitives that we need. 
%
% Particularly, 
% its encryption algorithm can encrypt a vector of $n$ packed values
% into a ciphertext of a certain level $l_{max}$. 
% A ciphertext of level $l>1$ can be converted to level $l'$ where $0<l'<l$.
% The $\oplus$ and $\otimes$ operations must be performed on
% the ciphertexts of the same level. 
% An $\otimes$ or $CMult$ operation consumes one level;
% that is, the new ciphertext returned by such an operation 
% has a level that is one lower than the level of the ciphertext operand(s).   
% Note that, an operation involving higher-level ciphertexts
% has higher time complexity than that involving lower-level ones. 
% For the ciphertexts of the same level,
% the $\otimes$ operation has the highest time complexity 
% (i.e., it is the most time-consuming),
% the $Rot$ operation has a slightly lower complexity,
% and it is followed by the $CMult$ operation.
% The $\oplus$ has the lowest complexity. 

% In our proposed scheme, we use the following primitives in FHE:
% \begin{itemize}
%     \item 
%     $FHE\_encrypt(\vec{v})$ - FHE encryption of vector $\vec{v}$ of plaintexts.
%     \item 
%     $FHE\_decrypt(\vec{v})$ - FHE-decryption of vector $\vec{v}$ of ciphertexts.
%     \item
%     $FHE\_add(\vec{v}_0,\cdots,\vec{v}_{n-1})$ - additions of $n$ vectors of FHE ciphertext.
% \end{itemize}

\section{Proposed Framework}
\label{sec:overview}

\begin{figure}[htb]
    \centering
    \caption{Framework of Proposed Solution. 
    % Interactions illustrated in the figure: 
    % (I.1) Model Provider attests TEE and shares its secret key and the hyber-parameters of its outsourced CNN model with the TEE; 
    % (I.2) Based on the CNN model's architecture, the TEE randomly generates the keys for asymmetric leveled homomorphic encryption and securely sends the public key $pk$ to Model Provider;
    % (I.3) Model Provider uploads its model parameters encrypted with $pk$ to REE;
    % (I.4) TEE sends $pk$ and the evaluation key $evk$ to the REE;
    % (S.1) Data Provider attests TEE and shares its secret key with the TEE;
    % (S.2) TEE securely sends $pk$ to Model Provider;
    % (S.3) Data Provider uploads encoded and encrypted data to REE;
    % (S.4) REE sends inference result to TEE for decryption; 
    % (S.5) TEE decrypts the result and returns the result encrypted with Data Provider's secret key.
    }
    \label{fig:framework}
    \includegraphics[width=\linewidth]{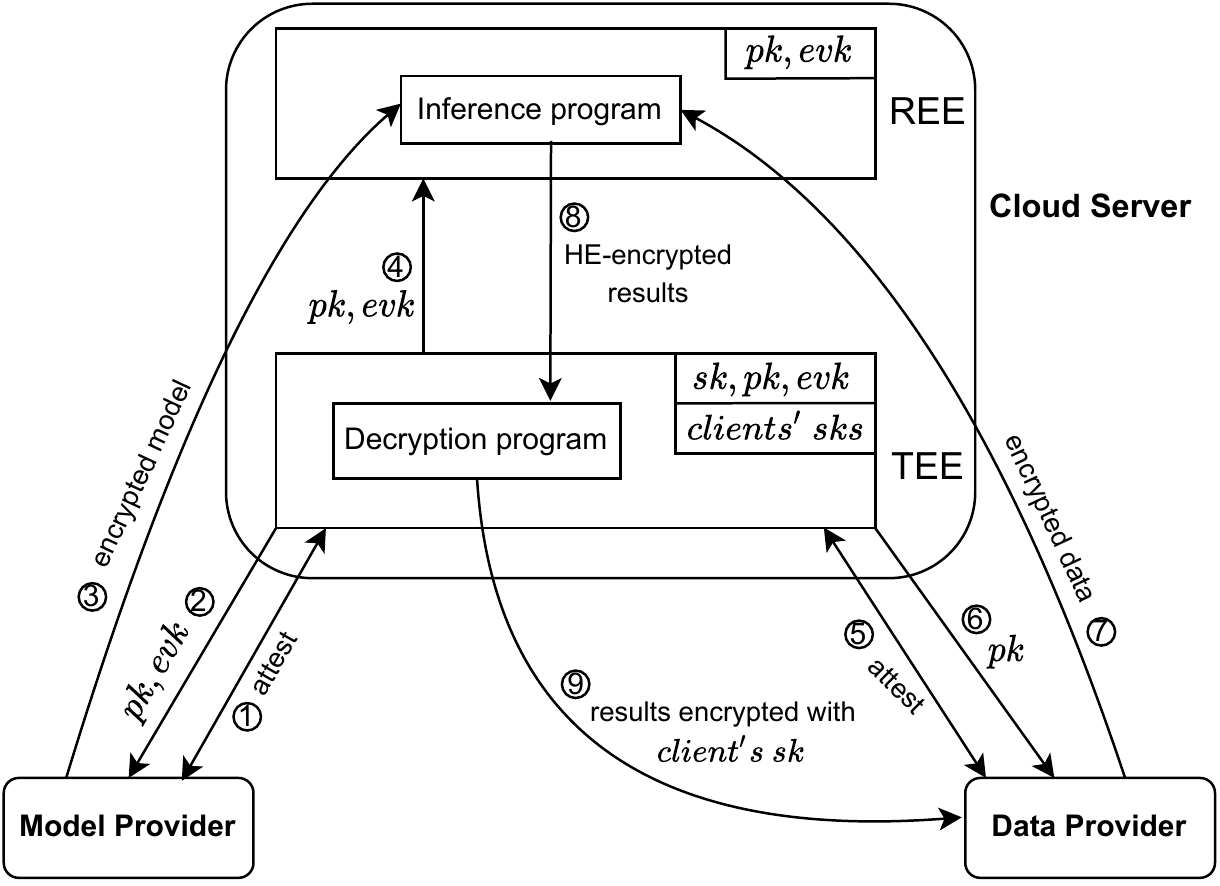}
\end{figure}

Figure~\ref{fig:framework} illustrates the framework of our proposed solution,
which is explained as follows. 
The framework involves four parties: 
the TEE of the cloud server, 
the REE of the cloud server,
the model provider, and the data provider. 
The interactions among these parties are as follows:
%
% \begin{enumerate}
%     \item
    \textcircled{1} The model provider attests the TEE.
    Once the attestation succeeds, it shares its secret key and 
    the hyper-parameters of its outsourced CNN model with the TEE.
    % \item
    \textcircled{2} Based on the CNN model's architecture, 
    the TEE calls algorithm $GenKey$ to randomly generate the keys for 
    asymmetric leveled homomorphic encryption and 
    securely sends the public key $pk$ to the model provider.
    % \item 
    \textcircled{3} The model provider encrypts its model parameters and 
    then uploads the encrypted parameters to the REE.
    % \item
    \textcircled{4} The TEE sends $pk$ and the evaluation key $evk$ to the REE.
    % \item 
    \textcircled{5} The data provider attests the TEE. 
    Once the attestation succeeds, it 
    shares its secret key with the TEE.
    % \item 
    \textcircled{6} TEE securely sends $pk$ to the data Provider.
    % \item 
    \textcircled{7} The data provider encodes and encrypts its data and 
    uploads the encrypted data to the REE.
    % \item 
    \textcircled{8} The REE conducts the inference based on encrypted model and encrypted data, and 
    sends the inference result to TEE for decryption.
    % \item 
    \textcircled{9} The TEE decrypts the result, and returns to the data provider 
    the result after it has been further encrypted with the data provider's secret key.
    % \item 
    \textcircled{10} The data provider decrypts the result with its secret key. 
% \end{enumerate}
%

Next, we elaborate on steps \textcircled{2}, \textcircled{3}, \textcircled{7} and \textcircled{8}, 
which form the LHE-based efficient inference scheme,
a performance-determining component of the proposed framework.

\section{LHE-based Inference Scheme}
\label{sec:scheme}

To support confidentiality-preserving inference with a CNN model 
of $c$ convolutional layers and $f$ fully connected layers,
the TEE first determines the highest encryption level ${\cal L}$ 
based on the topology of the model.
Specifically, if each layer consumes two encryption levels (as for the CNN model specified in Sec. \ref{subsec:cnn-model}),
it can set ${\cal L}=2(c+f)$. Next, the TEE 
initializes an LHE system by calling $KeyGen(2^\lambda,{\cal L})$ to obtain the keys; note that,
we set security parameter $\lambda$ to $128$ in our implementation. 
The public and evaluation keys are then distributed to the Server's REE and the clients,
while the private key is kept only by the TEE.

% as the highest encryption level and
% each ciphertext has ${\cal S}$ slots (i.e., it can encrypt up to ${\cal S}$ packed values).
% Here, ${\cal L}$ and ${\cal S}$ are derived from a certain security parameter $\lambda$,
% which is set to $128$ in our implementation. 

\subsection{Encoding and Encrypting Input Data and Filters}\label{sec:encoding}

In Steps \textcircled{3} and \textcircled{7} of the proposed framework,
the model provider should encode and encrypt its model parameters, 
and the data provider should encode and encrypt its input data,
before they upload them to the server.
%
% and the initial model parameters 
% (including the weights for filters and for fully-connected matrices), and
% then outsources them for the server to train the model. 
%
% The encoding and encryption should meet the following expectations:
% % \begin{itemize}
% %     \item 
%     after encoding and encryption, the data can be processed 
%     efficiently through the whole inference procedure, 
%     without decryption or re-encryption in the middle;
%     % \item
%     the slots available in a ciphertext should be used as much as possible 
%     for a high degree of parallel processing.
    %
    % \item
    % The ways that the weights are encoded and encrypted should facilitate efficient updating of them during the backward propagation.
    %
% \end{itemize}

% \subsubsection{Encoding and Encryption of Training Data}

\subsubsection{The Rationale}

To understand our proposed design, 
let us first briefly review the packing method in E2DM~\cite{jiang2018secure}.
We start with the simplest scenario where 
there is only one channel and one input matrix $I$ with $\beta_0^2$ values.
Following the notations introduced in Section~\ref{subsec:cnn-model},
let the only convolutional layer have kernel side $\gamma_0$ and stride $\delta_0$.
The method packs the $\beta_0^2$ values into $\gamma_0^2$ ciphertexts.
Denoting the ciphertexts as $\hat{C}_{i,j}$ for every $(i,j)\in\{0,\cdots,\gamma_0-1\}^2$,
each $\hat{C}_{i,j}$ encrypts the vector that packs all the values $I_{u,v}$ 
where $u-i$ and $v-j$ are multiples of stride $\delta_0$.
Accordingly, each element $F_{i,j}$ of each filter matrix $F$ is 
replicated to fill a vector and then encoded/encrypted to ciphertext $\hat{F}_{i,j}$; 
propagating through this layer results in a ciphertext $\sum_{0\leq i,j\leq\gamma_0-1}\hat{C}_{i,j}\otimes\hat{F}_{i,j}$. 
Note that, when there are multiple input matrices for a channel,
the above method can be straightly extended to have each ciphertext $\hat{C}_{i,j}$ 
pack and encrypt all the values $I_{u,v}$ from every input matrices $I$ 
where $u-i$ and $v-j$ are multiples of $\delta_0$.

% According to the notations introduced in Section~\ref{subsec:cnn-model}, 
% each input matrix is of $\beta_0^2$ elements.
% After one layer of filtering, it is reduced to an output matrix of $\beta_1^2$ elements,
% where $\beta_1 = 1 + \lfloor\frac{\beta_0-\gamma_0}{\delta_0}\rfloor$.
% To generalize, each convolutional layer $l$ reduces 
% an input matrix of $\beta_l^2$ elements to an output matrix of $\beta_{i+1}^2$ elements,
% where $\beta_{l+1}=1+\lfloor\frac{\beta_l-\gamma_l}{\delta_l}\rfloor$.

% \paragraph{Intuition}
To facilitate propagation through all the $c$ convolutional layers 
with packing/encryption only needed at the very beginning,
intuitively, we aim to encode/encrypt each input matrix 
as if there were only one {\em combined layer} 
that virtually includes all the $c$ layers in order.
We denote the parameters of the combined layer as:
\begin{itemize}
    \item 
    $\tilde{\delta}_0$ for {\em combined stride}, and
    \item 
    $\tilde{\gamma}_0$ for {\em combined kernel side} (i.e., $\tilde{\gamma}_0^2$ is the number of encrypted inputs for each channel).    
\end{itemize}
For convenience, we extend this notation and 
treat the combination of sequential layers from $l$ (where $0\leq l<c$) to $c-1$ as {\em combined layer from $l$};
thus, $\tilde{\delta}_l$ denotes {\em combined stride} and
$\tilde{\gamma}_l$ denotes {\em combined kernel side} for such a combined layer.

In our proposed design,
we aim to pack and encrypt data in a way that 
should meet the following requirements at each convolutional layer $l$:
First, $\tilde{\gamma}_l\geq\gamma_l$;
this is necessary as the input values should be 
multiplied with $\gamma_l^2$ elements of each filter respectively. 
Second, due to the rule of forward propagation through a convolutional layer, 
following relation should hold for 
the side $\tilde{\gamma}_l$ of each encrypted input matrix 
and the side of each output matrix (which is also the side $\tilde{\gamma}_{l+1}$
of each encrypted input matrix for layer $l+1$ when $l<c-1$):
\begin{equation}
    \tilde{\gamma}_{l+1} = 1 + \lfloor\frac{\tilde{\gamma}_l-\gamma_l}{\delta_l}\rfloor.
\end{equation}
Third, also due to the rule of the forward propagation,
following relation should hold between $\tilde{\delta}_l$
and $\tilde{\delta}_{l+1}$:
\begin{equation}
    \tilde{\delta}_l = \delta_l\cdot\tilde{\delta}_{l+1}.
\end{equation}
Particularly for the last convolutional layer, 
% \begin{equation}
    $\tilde{\delta}_{c-1} = \delta_{c-1}$.
% \end{equation}
%
Setting $\tilde{\gamma}_{c-1}=\gamma_{c-1}$ and guided by the above requirements,
we can derive the following general settings of the parameters for each layer $l$:
\begin{equation}
    \tilde{\gamma}_l = 1 + \sum_{i=l}^{c-1}(\gamma_i-1)\prod_{j=0}^{i-1}\delta_j,
\end{equation}
and 
\begin{equation}
    \tilde{\delta}_l = \prod_{i=l}^{c-1}\delta_i.
\end{equation}
Based on the above parameter settings, 
we elaborate our proposed method for encoding and encryption in the following. 

% the stride is the product of all the layers' strides; that is,  
% \begin{equation}
%     \tilde{\delta}_0 = \prod_{i=0}^{c-1}\delta_i,
% \end{equation}
% which we call {\em combined stride}.
% Hence, the corresponding {\em combined kernel} has side $\tilde{\gamma}_0$ 
% where 
% \begin{equation}
%     \tilde{\gamma}_0 = 1 + \sum_{i=0}^{c-1}(\gamma_i-1)\prod_{j=0}^{i-1}\delta_j,
% \end{equation}
% and accordingly, each {\em encoded input matrix} (i.e., {\em combined matrix}) 
% has side $\tilde{\beta}_0$ 
% where
% \begin{equation}
%     \tilde{\beta}_0 = 1 + \lfloor\frac{\beta_0 - \tilde{\gamma}_0}{\tilde{\delta}_0}\rfloor.
% \end{equation}

\subsubsection{Encoding and Encrypting Input Data}

For each channel $i$, 
all of its $n$ input matrices are encoded as follows.
Let these matrices be denoted as $I_i^{(j)}$ for every $j\in\{0,\cdots,n-1\}$, and 
each value of $I_i^{(j)}$ be denoted as $I^{(j)}_{i,x,y}$ for $x,y\in\{0,\cdots,\beta_0-1\}$.
All the values are encoded/encrypted into $\tilde{\gamma}_0^2$ ciphertexts,
each indexed by $(i,u,v)$ for $(u,v)\in\{0,\cdots,\tilde{\gamma}_0-1\}^2$.
Each ciphertext $\hat{C}^{(0)}_{i,u,v}$ encodes/encrypts the following vector:

\begin{equation}
\resizebox{\linewidth}{!}{
$\displaystyle
    \begin{pmatrix}\label{eq:initial-input-format-1}
        \vec{I}_{i,u,v} & \vec{I}_{i,u,v+\tilde{\delta}_0} & \cdots & \vec{I}_{i,u,v+(\tilde{\beta}_0-1)\tilde{\delta}_0} \\
        \vec{I}_{i,u+\tilde{\delta}_0,v} & 
        \vec{I}_{i,u+\tilde{\delta}_0,v+\tilde{\delta}_0} 
        & \cdots 
        & \vec{I}_{i,u+\tilde{\delta}_0,v+(\tilde{\beta}_0-1)\tilde{\delta}_0} \\
        \vdots & \vdots & \ddots & \vdots \\
        \vec{I}_{i,u+(\tilde{\beta}_0-1)\tilde{\delta}_0,v} & 
        \vec{I}_{i,u+(\tilde{\beta}_0-1)\tilde{\delta}_0,v+\tilde{\delta}_0} & 
        \cdots & 
        \vec{I}_{i,u+(\tilde{\beta}_0-1)\tilde{\delta}_0,v+(\tilde{\beta}_0-1)\tilde{\delta}_0}
    \end{pmatrix},
$
}
\end{equation}
where for each $(x,y)\in\{0,\cdots,\tilde{\gamma}_0-1\}^2$,
\begin{equation}\label{eq:initial-input-format-2}
    \vec{I}_{i,x,y} = ({I}_{i,x,y}^{(0)},~\cdots,~{I}_{i,x,y}^{(n-1)}).
\end{equation}

\subsubsection{Encoding and Encrypting Filters}

Due to the above method for encoding/encrypting input data,
all the values encoded/encrypted in one same ciphertext needs to be multiplied with one same element of a filter at a time.
Hence, each element $F_{i,j}$ of a filter $F$ should be duplicated for $n\cdot\tilde{\beta}_0^2$ times 
and then encoded/encrypted into a ciphertext denoted as $\hat{F}_{i,j}$.

\subsection{Propagation through Convolutional Layer $l$}

% To facilitate the description, 
% we first generalize the definitions of 
% %$\tilde{\delta}_0$, 
% $\tilde{\gamma}_0$ 
% % and $\tilde{\beta}_0$ 
% to those of
% % $\tilde{\delta}_l$, 
% $\tilde{\gamma}_l$ 
% % and $\tilde{\beta}_l$ 
% for $l=0,\cdots,c-1$, 
% as follows:
% % \begin{equation}\label{eq:tilde-delta-l}
% %     \tilde{\delta}_l = \prod_{i=l}^{c-1}\delta_i,
% % \end{equation}
% %
% \begin{equation}
%     \tilde{\gamma}_l = 1 + \sum_{i=l}^{c-1}(\gamma_i-1)\prod_{j=l}^{i-1}\delta_j,
% \end{equation}
% %
% % and 
% % \begin{equation}
% %     \tilde{\beta}_l = 1 + \lfloor\frac{\beta_l-\tilde{\gamma}_l}{\tilde{\delta}_l}\rfloor.
% % \end{equation}
% which is equivalent to 
% \begin{equation}
%     \tilde{\gamma}_l = \gamma_l + (\tilde{\gamma}_{l+1}-1)\cdot \delta_l.
% \end{equation}

With the above method for encoding and encryption,
each propagation layer $l$ has $\tilde{\gamma}_l^2$ input ciphertexts for each channel;
they can be treated as elements of an encrypted input matrix of dimensions $\tilde{\gamma}_l\times\tilde{\gamma}_l$.
We denote these input ciphertexts as $\hat{C}^{(l)}_{i,u,v}$ for 
every channel $i\in\{0,\cdots,\alpha_{l}-1\}$ and every pair $(u,v)\in\{0,\cdots,\tilde{\gamma}_l-1\}^2$.
Convolutional operations are then performed between 
these input ciphertexts and the encrypted elements of the filters.
The encrypted elements of the filters are denoted as 
$\hat{F}^{(l,k)}_{i,x,y}$ for 
every channel $i$,
every filter with index $k\in\{0,\cdots,\epsilon_l-1\}$ of the channel, 
and every element index $(x,y)\in\{0,\cdots,\gamma_l-1\}^2$ of a filter matrix.
Then, the activation function, which is a square function in this paper, 
is applied at each of the resulting neurons. 
Finally, propagating through the layer results in the following outputs:
$\hat{C}^{(l+1)}_{k,u,v}$ for every output channel $k\in\{0,\cdots,\epsilon_l-1\}$
and every pair $(u,v)\in\{0,\cdots,\tilde{\gamma}_{l+1}-1\}^2$.
Note that, the outputs become the inputs of the next layer;
when $l=c=1$, each output channel has only one output ciphertext, 
i.e., $\tilde{\gamma}_{c}=1$.
The procedure is formally presented in Algorithm~\ref{alg:fwd-conv-layer}.

% Right before the propagation through a convolutional layer $l$,
% for each channel $i$, 
% each input matrix should have already encrypted into $\tilde{\gamma}_l^2$ ciphertexts,
% each of which can be indexed by a unique $(u,v)\in\{0,\cdots,\tilde{\gamma}_l-1\}^2$;
% we denote the afore-mentioned input matrix as $\vec{I}^{(l)}_{i,u,v}$ and 
% the ciphertext as $\vec{C}^{(l)}_{i,u,v}$. 
% %
% For the same channel $i$,
% each element with index $(u,v)\in\{0,\cdots,\gamma_l-1\}^2$ 
% of each filter matrix with index $k\in\{0,\cdots,f_l-1\}$ 
% is encrypted to ciphertext $\hat{F}^{(l,k)}_{i,u,v}$.

% While propagating through the layer,

% Algorithm~\ref{alg:fwd-conv-layer} formally presents how the forward propagation is conducted
% through a convolutional layer $l$.

\begin{algorithm}[htb]
\SetAlgoLined
\caption{Propagation through Conv. Layer $l$}
\label{alg:fwd-conv-layer}

% \begin{algorithmic}[1]

\For{$k\in\{0,\cdots,\epsilon_l-1\}$}{ 
        \For{$(u,~v)\in\{0,\cdots,\tilde{\gamma}_{l+1}-1\}^2$}{
            % \For{$v\in\{0,\cdots,\tilde{\gamma}_{l+1}-1\}$} 
                $\hat{C}^{(l+1)}_{k,u,v}\leftarrow 0$; 
                \Comment{initialize each output}\;
                $(u',~v')\leftarrow (\delta_l\cdot u,~\delta_l\cdot v)$\;
                % \State $v'\leftarrow \delta_l\cdot v$
                \For{$(x,~y)\in\{0,\cdots,\gamma_l-1\}^2$}{
                    % \For{$y\in\{0,\cdots,\gamma_l-1\}$}
                        \For{$i\in\{0,\cdots,\alpha_l-1\}$}{ 
                            $\hat{C}^{(l+1)}_{k,u,v}\oplus = \hat{C}^{(l)}_{i,u'+x,v'+y}\otimes\hat{F}^{(l,k)}_{i,x,y}$;
                        }
                        % \EndFor
                    % \EndFor
                }
                % \EndFor 
                $\hat{C}^{(l+1)}_{k,u,v}\leftarrow \hat{C}^{(l+1)}_{k,u,v}\otimes\hat{C}^{(l+1)}_{k,u,v}$; 
                \Comment{activation}
            % \EndFor
        }
        % \EndFor
}
% \EndFor

% \end{algorithmic}
\end{algorithm}

% \subsection{Propagation through Pooling Layer $l$}

% (skip for now)

\subsection{Propagation through Fully-connected Layer $l$}

% We use the following notations to facilitate our presentation. 
Recall that the input to layer $l$ is 
the intermediate results of processing $n$ original inputs,
and the processing for these $n$ inputs are in parallel independently.
%
% We use $\iota_l$ to denote the number of values in the input to layer $l$ that are from each of the $n$ original inputs;
% hence, totally there are $\iota_l\cdot n$ distinct input values encoded/encrypted in the input to layer $l$.
%
Due to our method for encoding (as in Section~\ref{sec:encoding}),
the input values that are of the same offset but from $n$ different original inputs 
are encoded consecutively (as in Eq.~(\ref{eq:initial-input-format-2})),
we call such sequence of $n$ values as a {\em parallel input set (pi-set)}. 
%Obviously, each set encodes $n$ values. 
We use $\iota_l$ to denote the number of pi-sets encoded/encrypted into the input of layer $l$;
thus, the input of layer $l$ encodes/encrypts totally $\iota_l\cdot n$ values. 
% Hence, the input to layer $l$ encodes/encrypts of $\iota_l$ sets (i.e., totally $\iota_l\cdot n$) of distinct input values.
%
We further use $\iota'_l$ to denote the number of ciphertexts in the input to layer $l$, and denote
the ciphertexts as $\hat{C}^{(l)}_0,\cdots,\hat{C}^{(l)}_{\iota'_l-1}$.
Our design evenly distributes the $\iota_l$ pi-sets to the $\iota'_l$ ciphertexts; thus, 
each ciphertext encodes/encrypts $\iota''_l=\frac{\iota_l}{\iota'_l}$ pi-sets.
We also assume $\iota''_l$ is an integer for the convenience of presentation. 

The input to layer $l$ has two types.
% \begin{itemize}
%     \item 
    Type I (ciphertext input without replication):
    % For this type, 
    % such as the resulting ciphertexts from the afore-discussed convolutional layers, 
    each ciphertext encodes/encrypts multiple pi-sets in the format similar to 
    Eq.~(\ref{eq:initial-input-format-1}) and (\ref{eq:initial-input-format-2}), 
    in which the pi-sets are not replicated.
    % \item 
    Type II (ciphertext input with replication):
    each ciphertext contains one pi-set which is replicated for multiple times. 
% \end{itemize}
Next, we present the algorithms for propagating through layer $l$ with each type of input.

\subsubsection{Propagation with Type I Input}
\label{sec:forward-full-1}

%forward propagation needs rotation; but backward does not

% Generally, the convolutional layers result in a setting where 
% each input ciphertext contains multiple sets of input elements. 

% Suppose all the $\iota_l$ sets of input elements are stored in $\iota'_l$ ciphtertexts,
% which are denoted as $\tilde{C}^{(l)}_0,\cdots,\tilde{C}^{(l)}_{\iota'_l-1}$,
% and each of such ciphertext contains $\iota''_l$ sets of input elements; 
% hence, $\iota_l=\iota'_l\cdot\iota''_l$.
% Note that, the size of each set is $n$ as there are $n$ sets of learning data processed per epoch.

In this case, each input ciphertext $\hat{C}^{(l)}_j$ for $j\in\{0,\cdots,\iota'_l-1\}$
encodes and encrypts $\iota''_l$ pi-sets.
As presented in Section~\ref{subsec:cnn-model},
the weight matrix for the layer is $M^{(l)}$ with dimensions $\iota_l\times o_l$;
let us denote the elements of $M^{(l)}$ as 
$M^{(l)}_{u,v}$ where $u\in\{0,\cdots,o_l-1\}$ and $v\in\{0,\cdots,\iota_l-1\}$.  
To facilitate efficient forward propagation through this layer,
the elements of $M^{(l)}$ are encoded and encrypted to $o_l\cdot\iota'_l$ ciphertexts,
denoted as $\hat{M}^{(l)}_{i,j}$ for every $i\in\{0,\cdots,o_l-1\}$ and every $j\in\{0,\cdots,\iota'_l-1\}$.
Here, each $\hat{M}^{(l)}_{i,j}$ encrypts the following vector
\begin{equation}
    (\vec{M}^{(l)}_{i,j\cdot\iota''_l}, \cdots, \vec{M}^{(l)}_{i,(j+1)\cdot\iota''_l-1}),
\end{equation}
where for each $w\in\{j\cdot\iota''_l,\cdots,(j+1)\cdot\iota''_l-1\}$,
$\vec{M}^{(l)}_{i,w}$ stands for a sequence of $n$ duplicated values of ${M}^{(l)}_{i,w}$; i.e., 
\begin{equation}
    \vec{M}^{(l)}_{i,w} = ({M}^{(l)}_{i,w},\cdots,{M}^{(l)}_{i,w}). 
\end{equation}
% $M_l[i][j\cdot\iota''_l],\cdots,M_l[i][(j+1)\cdot\iota''_l-1]$.

Algorithm~\ref{alg:fwd-full-layer-I} formally presents the forward propagation.
%
% To understand this algorithm,
% let us consider how the $i$=th output values for all the $n$ original inputs are computed.
% We denote the 
% If there is no encryption, this output value should be computed as 
% \begin{equation}
%     \sum_{j=0}^{\iota_l-1} M^{(l)}_{i,j}\cdot 
% \end{equation}
%
% Specifically, each output ciphertext $\hat{C}^{(l+1)}_i$ for 
% $i\in\{0,\cdots,o_l-1\}$ is computed as follows.
% First, with the loop on lines 3-5, 
%
% is conducted through a type-I fully-connected layer $l$.
This results in $o_l$ ciphertexts, each encoding/encrypting 
$\frac{\cal S}{n}$ replicas of a set of $n$ output values.
Note that, if the output is used as the input of the next layer,
it becomes a type II input.

\begin{algorithm}[htb]
\SetAlgoLined
\caption{Propagation through Fully-connected Layer $l$ (with Type I Input)}
\label{alg:fwd-full-layer-I}

% \begin{algorithmic}[1]

\For{$i\in\{0,\cdots,o_l-1\}$}{
    $\hat{C}^{(l+1)}_i\leftarrow 0$; \Comment{initialize each output ciphertext}\;
    \For{$j\in\{0,\cdots,\iota'_l-1\}$}{
        $\hat{C}^{(l+1)}_i \oplus = \hat{C}^{(l)}_{j}\otimes\hat{M}^{(l)}_{i,j}$;
    }
    \For{$j\in\{1,\cdots,\log(\frac{\cal S}{n})\}$}{
        $\hat{C}^{(l+1)}_i \oplus= Rot(\hat{C}^{(l+1)}_i,j\cdot n)$;
    }
    $\hat{C}^{(l+1)}_i\leftarrow \hat{C}^{(l+1)}_i\otimes\hat{C}^{(l+1)}_i$; 
    \Comment{activation}
}
    % \State $\tilde{C}^{(l+1)}_i\leftarrow 0$ \Comment{each output element $i$}
    % \For{$j\in\{0,\cdots,\iota'_l-1\}$} 
    %     \State $\tilde{C}^{(l+1)}_i += \tilde{C}^{(l)}_{j}\bigodot\tilde{M}^{(l)}_{i,j}$
    % \EndFor
    % \For{$j\in\{1,\cdots,\log(\frac{s}{n})\}$}
    %     \State $\tilde{C}^{(l+1)}_i += Rotate(\tilde{C}^{(l+1)}_i,j\cdot n)$
    % \EndFor
% \EndFor

% \end{algorithmic}
\end{algorithm}

\subsubsection{Propagation with Type II Input}

%forward propagation or backward propagation does not need rotation

In this case,
the input has $\iota'_l=\iota_l$ ciphertexts
each encrypting $\frac{\cal S}{n}$ replicas of only one input set (i.e., $n$ input values from $n$ original inputs).
To facilitate the propagation,
the elements of weight matrix $M^{(l)}$ are encrypted to 
$\iota_l\cdot\lceil\frac{o_l\cdot n}{\cal S}\rceil$ ciphtertexts,
denoted as $\hat{M}^{(l)}_{i,j}$ for 
$i\in\{0,\cdots,\iota_l-1\}$ and $j\in\{0,\cdots,\lceil\frac{o_l\cdot n}{\cal S}\rceil-1\}$,
where each $\hat{M}^{(l)}_{i,j}$ encodes and encrypts the following vector
\begin{equation}
    (\vec{M}^{(l)}_{j\cdot\frac{\cal S}{n},i},\cdots,\vec{M}^{(l)}_{\min\{(j+1)\cdot\frac{\cal S}{n},o_l\}-1,i}),
\end{equation}
where for each $w\in\{j\cdot\frac{\cal S}{n},\cdots,\min\{(j+1)\cdot\frac{\cal S}{n},o_l\}-1\}$, 
$\vec{M}^{(l)}_{w,i}$ stands for a sequence of $n$ duplicated values of ${M}^{(l)}_{w,i}$.

% weights
% $M_l[j\cdot\frac{s}{n}][i],\cdots,M_l[\min\{j\cdot\frac{s}{n},o_l\}-1][i]$.

Algorithm~\ref{alg:fwd-full-layer-II} formally presents how the forward propagation
is conducted through a type-II fully-connected layer $l$.
This results in $\lceil\frac{o_l\cdot n}{\cal S}\rceil$ ciphertexts
each encoding/encrypting $\frac{\cal S}{n}$ pi-sets.
Note that, if the output is used as the input for the next layer,
it becomes a Type I input. 

\begin{algorithm}[htb]
\SetAlgoLined
\caption{Propagation through Fully-connected Layer $l$ (with Type II Input)}
\label{alg:fwd-full-layer-II}

\For{$i\in\{0,\cdots,\lceil\frac{o_l\cdot n}{\cal S}\rceil-1\}$}{
    $\hat{C}^{(l+1)}_i\leftarrow 0$; \Comment{each output element $i$}\;
    \For{$j\in\{0,\cdots,\iota_l-1\}$}{
        $\hat{C}^{(l+1)}_i \oplus= \hat{C}^{(l)}_{j}\otimes\hat{M}^{(l)}_{i,j}$
    }
    $\hat{C}^{(l+1)}_i\leftarrow \hat{C}^{(l+1)}_i\otimes\hat{C}^{(l+1)}_i$; 
    \Comment{activation}
}

\end{algorithm}

\begin{figure*}[htb]
    \centering
    \caption{An Example.}
    % of Proposed Solution. Each matrix represents the original data and each vector represents one ciphertext with multiple data packed.}
    \label{fig:example}
    \includegraphics[width=\textwidth]{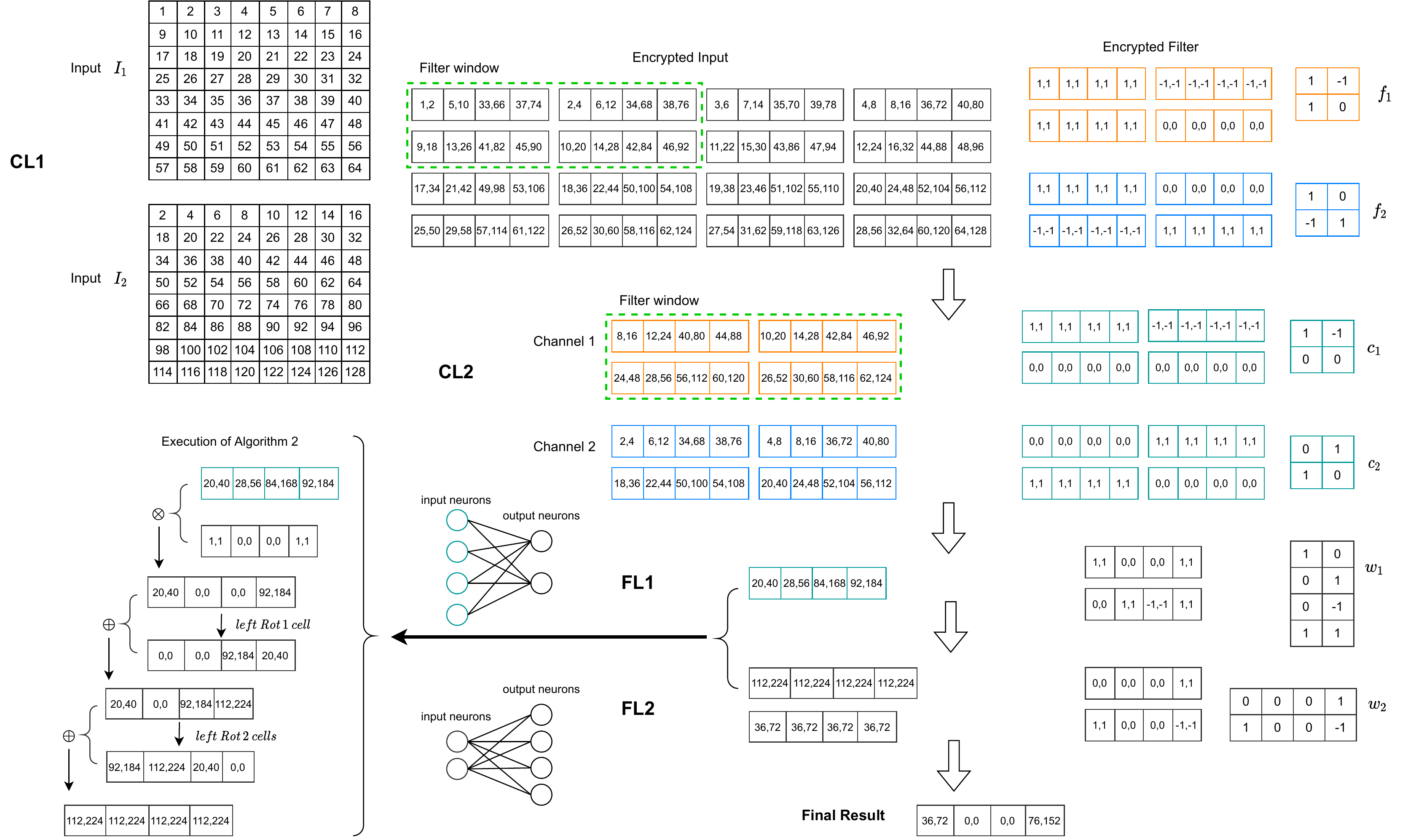}
\end{figure*}

\subsection{An Example}

Figure~\ref{fig:example} illustrates a simple example,
where our proposed scheme is applied to a CNN model of $2$ CLs and $2$ FLs and 
activation function is not considered.
Two $8\times 8$ input matrices, 
labelled as $I_1$ and $I_2$, 
are processed in parallel. 
Thus, $16$ ciphertexts are constructed to encode/encrypt the values of the input matrices.
Each ciphertext encodes/encrypts $8$ input values,
represented as $4$ cells.
Each cell contains $2$ values from the $2$ input matrices respectively; recall that
such a cell is called pi-set in the description of our scheme. 

CL1 has two $2\times2$ filter matrices, 
labelled as $f_1$ and $f_2$, each has one channel. 
CL2 also has two $2\times2$ filters, 
labelled as $c_1$ and $c_2$, each belonging to a separated channel. 
Each filter is encrypted to $4$ ciphertexts, 
and each ciphtertext contains $8$ replicated elements to 
match the ciphertexts for input values.
The output from the two CLs is one single ciphertext,
which is a Type I input to layer FL1. 

FL1 has $4$ input and $2$ output neurons as inferred from its $4\times2$ weight matrix.
The elements of the matrix is encrypted to $2$ ciphertexts 
corresponding to the $2$ output neurons respectively.
Now, the outputs should be computed as the dot-products of 
the vector encoded/encrypted in the input ciphertext and 
each of the vectors encoded/encrypted in the weight-matrix's ciphertexts. 
This is accomplished according to Algorithm~\ref{alg:fwd-full-layer-I},
and the call-out on the left-hand side illustrates 
the steps for computing the first of the outputs: 
% \begin{itemize}
%     \item 
    
{\underline{First}}, the input ciphertext that encrypts $(20,40,28,56,84,168,92,184)$ HE-multiplies with
    the first ciphertext of the weight matrix, which encrypts $((1,1,0,0,0,0,1,1)$;
    this results in the ciphertext that encrypts $(20,40,0,0,0,0,92,184)$.
    
    % \item 
{\underline{Second}}, the ciphertext for $(20,40,0,0,0,0,92,184)$ HE-adds 
    a copy of itself which but also rotates (to left) for one cell,
    and this results in a ciphertext for $(20,40,0,0,92,184,112,224)$.
    
    % \item
{\underline{Third}}, the ciphertext for HE-adds a copy of itself which also rotates (to left) for $2$ cells,
    and this results in a ciphertext for $(112,224,112,224,112,224,112,224)$.

% \end{itemize}
The result of the above steps becomes a type II input for the next layer, FL2.
For FL2, the weights connected to the same input neuron are packed into one ciphertext;
hence, there are $2$ ciphertexts for the weights.
Each of the $2$ ciphertexts multiplies with the corresponding input ciphertext,
and resulting $2$ product-ciphertexts are added up to get the final result,
which encrypts $(36,72,0,0,0,0,76,152)$.

% For filter $f_1$ and $f_2$, each is encrypted into 4 ciphertexts.
% The outputs of CL1 has 2 channels and each channel contains 4 ciphertexts.
% There are 1 filter in CL2 and contains 2 channels $c_1$ and $c_2$, each channel is of dimension $2\times2$.
% Same as filters in CL1, each value inside filter is duplicated and encrypted into one ciphertext.
% After 2 CLs, the outputs are packed into one ciphertext.
% The weight matrix $w_1$ in FL1 and $w_2$ in FL2 has dimension $4\times2$ and $2\times4$ respectively.
% $w_1$ has all weights connected to the same output neuron packed together. 
% Then it requires 2 ciphertext to pack $w_1$.
% In order to get the correct results of FL1, we need to rotate the results after the input and packed weight multiplication.
% The detailed steps for the rotation to get the FL1 output
% (i.e., $(76,152,76,152,76,152,76,152)$) is also described in figure~\ref{fig:example}.
% In FL2, the weights connected to the same input neuron are packed into one ciphertext.
% Then $w_2$ is packed into 2 ciphertexts.

\section{Optimizations on Packing}
\label{sec:optimize}

With the above scheme,
each initial input ciphertext encodes/encrypts $n\cdot\tilde{\beta}_0^2$ input values.
If $n\cdot\tilde{\beta}_0^2$ is much less than ${\cal S}$ (i.e., the number of available slots in a ciphertext),
the slots will not be effectively utilized and the computation/communication efficiency will be negatively affected.
To address this limitation, we propose cross-channel and cross-filter packing in this section.
%
% When the number of available data sets $n$ is small or the slots number $s$ is much larger than $n\cdot\tilde{\beta}_0^2$, in order to fully utilize the slots in a ciphertext, we propose two ways to optimize the packing methods of the convolutional layer from the basic scheme. 
%
Specifically, 
letting $r$ be the largest power of two which is no greater than $\frac{\cal S}{n\cdot\tilde{\beta}_0^2}$, 
the inputs from $r$ channels can be encoded together such that more input values can be processed in parallel, 
or each ciphertext can encode/encrypt $r$ replicas of the original $n\cdot\tilde{\beta}_0^2$ values such that
the same input can be processed with multiple filters in parallel. 
Such changes, however, also demand according changes to the procedure of propagation through each convolutional layer.
In this section, we elaborate on these optimizations. 

%In this section, we elaborate the methods to parallel channels and filters in convolutional layer $l$.

\subsection{Cross-Channel Packing}

With cross-channel packing,
input values from $r$ different channels can be encoded/encrypted 
% Multiple ciphertexts of different channels can be combined 
into one ciphertext; accordingly,
the elements of the filters associated with these channels should be encoded together accordingly.  
% the corresponding ciphertexts for filter matrix elements should be combined accordingly.
Such a design reduces the number of input ciphertexts at convolutional layer $l$ to 
$\tilde{\gamma}_l^2\lceil\frac{\alpha_l}{r} \rceil$ and reduces  
the number of ciphertexts that encode/encrypt filters to $\epsilon_l\lceil\frac{\alpha_l}{r} \rceil\beta_l^2$.

More specifically,  
all the $\alpha_l$ channels at layer $l$ are grouped into $\lceil\frac{\alpha_l}{r}\rceil$ groups,
where each channel group $i'$ includes channels $\{i'\cdot r,\cdots,\min\{(i'+1)\cdot r,\alpha_0\}-1\}$.
All the plaintext input values to this layer are now encoded/encrypted into ciphertexts 
$\vec{C}^{(l)}_{i',u,v}$ for every $i'\in\{0,\cdots,\lceil\frac{\alpha_l}{r}\rceil-1\}$
and every $(u,v)\in\{0,\cdots,\tilde{\gamma}_l-1\}^2$.
Here, each $\vec{C}^{(l)}_{i',u,v}$ encodes/encrypts the concatenation of $r$ vectors formatted as as 
Eq.~(\ref{eq:initial-input-format-1}) and (\ref{eq:initial-input-format-2})
for every $i\in\{i'\cdot r,\cdots,\min\{(i'+1)\cdot r,\alpha_l\}-1\}$.
Accordingly, for every $i'$, every $k\in\{0,\epsilon_l-1\}$ and every $(x,y)\in\{0,\cdots,\beta_l-1\}^2$,
a ciphertext $\hat{F}^{(l,k)}_{i',x,y}$ is constructed to encode/encrypt 
the elements with index $(x,y)$, which is replicated for $n$ copies, of the $k$-th filers in every channels of channel group $i'$.

Based on the above encryption strategy,
the procedure of propagation through this layer is formalized in Algorithm~\ref{alg:fwd-conv-layer-cross-channels}.
As we can see, most part of this algorithm is similar to Algorithm~\ref{alg:fwd-conv-layer}, but it
reduces the iteration number from $\alpha_l$ to $\lceil\frac{\alpha_l}{r}\rceil$ in the innermost layer of loop (lines 6-9).
The algorithm further employs rotation to sum up subsets of the elements encoded/encrypted in each ciphertext,
which is similar to Algorithm~\ref{alg:fwd-full-layer-I}.
Each ciphertext output by this algorithm contains $r$ replicas of data set encoded/encrypted in the cross-filter pattern.
If the output is used as the input of next convolutional layer,
they can be processed as described next.

\begin{algorithm}[htb]
\SetAlgoLined
\caption{Propagation Through Convolutional Layer $l$ with Cross-Channel Packing}
\label{alg:fwd-conv-layer-cross-channels}

\For{$k\in\{0,\cdots,\epsilon_l-1\}$}{ %\Comment{$\epsilon_l=\alpha_{l+1}$ each output channel}
    \For{$(u,v)\in\{0,\cdots,\tilde{\gamma}_{l+1}-1\}^2$}{
        $\hat{C}^{(l+1)}_{k,u,v}\leftarrow 0$; \Comment{each element of output matrix}\;
        $(u',~v')\leftarrow (\delta_l\cdot u,~\delta_l\cdot v)$\;
        \For{$(x,y)\in\{0,\cdots,\gamma_l-1\}^2$}{
            \For{$i'\in\{0,\cdots,\lceil\frac{\alpha_l}{r}\rceil-1\}$}{ 
                \Comment{iterating each channel group}\;
                $\hat{C}^{(l+1)}_{k,u,v} \oplus= \hat{C}^{(l)}_{i',u'+x,v'+y}\otimes \hat{F}^{(l,k)}_{i',x,y}$;
            }
        }
        \For{$j\in\{1,\cdots,\log(r)\}$}{ %\Comment{sum up all channels}
                $\tilde{C}^{(l+1)}_{k,u,v} += Rot(\tilde{C}^{(l+1)}_{k,u,v},~j\cdot n\cdot\tilde{\beta}_0^2)$;
        }
        $\hat{C}^{(l+1)}_{k,u,v}\leftarrow \hat{C}^{(l+1)}_{k,u,v}\otimes\hat{C}^{(l+1)}_{k,u,v}$; 
                \Comment{activation}
    }
}

\end{algorithm}

% \begin{algorithm}[htb]
% \caption{Propagation Through Convolutional Layer $l$ with Cross Channels Packing}
% \label{alg:fwd-conv-layer-cross-channels}

% \begin{algorithmic}[1]

% \For{$k\in\{0,\cdots,\epsilon_l-1\}$} \Comment{$\epsilon_l=\alpha_{l+1}$ each output channel}
%     \For{$u\in\{0,\cdots,\tilde{\gamma}_{l+1}-1\}$}
%         \For{$v\in\{0,\cdots,\tilde{\gamma}_{l+1}-1\}$} 
%             \State $\tilde{C}^{(l+1)}_{k,u,v}\leftarrow 0$ \Comment{each element of output matrix}
%             \State $u'\leftarrow \delta_l\cdot u$
%             \State $v'\leftarrow \delta_l\cdot v$
%             \For{$x\in\{0,\cdots,\gamma_l-1\}$}
%                 \For{$y\in\{0,\cdots,\gamma_l-1\}$}
%                     \For{$i\in\{0,\cdots,\lceil\frac{\alpha_l}{r}\rceil-1\}$} \Comment{each cross channels packing}
%                     \State $\tilde{C}^{(l+1)}_{k,u,v} += \tilde{C}^{(l)}_{i,u'+x,v'+y}\bigodot \hat{F}^{(l,k)}_{i,x,y}$
%                     \EndFor
%                 \EndFor
%             \EndFor
%             \For{$j\in\{1,\cdots,\log(r)\}$} \Comment{sum up all channels}
%                 \State $\tilde{C}^{(l+1)}_{k,u,v} += Rotate(\tilde{C}^{(l+1)}_{k,u,v},j\cdot n\cdot\tilde{\beta}_0^2)$
%             \EndFor
%         \EndFor
%     \EndFor
% \EndFor

% \end{algorithmic}
% \end{algorithm}

\subsection{Cross-Filter Packing}

With cross-filter packing,
distinct input values are packed as pi-sets and then all these distinct values 
as whole is replicated for $r$ times.
Specifically for a layer $l$ with cross-filter packing input,
the input includes $\alpha_l\cdot\tilde{\gamma}_l^2$ ciphertexts. 
Each ciphertext $\hat{C}^{(l)}_{i,u,v}$ for channel $i\in\{0,\cdots,\alpha_l-1\}$
and $(u,v)\in\{0,\cdots,\tilde{\gamma}_l-1\}^2$ encodes/encrypts 
$r$ replicas of the $\tilde{\beta}_l$ pi-sets 
that are formatted similarly to Eq.~(\ref{eq:initial-input-format-1}) and (\ref{eq:initial-input-format-2}).

The existence of replication in the ciphertexts provides an opportunity for 
each ciphertext to be multiplied with a ciphertext encoding/encrypting elements from multiple filters at a time;
this way, increased level of parallel processing can be attained. 
Specifically, for each channel $i$, all the $\epsilon_l$ filter matrices are grouped into $\lceil\frac{\epsilon_l}{r}\rceil$ groups
such that each group $k'\in\{0,\cdots,\lceil\frac{\epsilon_l}{r}\rceil-1\}$ contains 
filters $\vec{F}^{(l,k)}$ for every $k\in\{k'\cdots r, \cdots, \min\{(k+1)\cdot r, \epsilon_l\}-1\}$. 
Then, for every such filter group $k'$ and every pair $(x,y)\in\{0,\cdots,\beta_l-1\}^2$,
a ciphertext $\hat{F}^{(l,k')}_{i,x,y}$ is constructed to encode/encrypt every element with index $(x,y)$, 
which is replicated for $n$ copies, of every filters in group $k'$ of channel $i$.

Based on the above encoding/encryption strategy, 
the propagation through this layer is formalized in Algorithm~\ref{alg:fwd-conv-layer-cross-filters}. 
This algorithm is similar to Algorithm~\ref{alg:fwd-conv-layer}, 
except for that it iterates through filter groups (each containing up to $r$ filters) instead of individual filters;
this way, we can gain a speed-up up to a factor of $r$. 
Interestingly,
the output ciphertexts, when used as input for next layer,
become cross-channel packed,
which can thus be processed as described in the previous subsection.

\begin{algorithm}[htb]
\SetAlgoLined
\caption{Propagation Through Convolutional Layer $l$ with Cross-Filter Packing}
\label{alg:fwd-conv-layer-cross-filters}

% \For{$k\in\{0,\cdots,\lceil\frac{\epsilon_l}{r}\rceil-1\}$}{ %\Comment{each cross filters packing}
%   Same as Lines $ 2-10$ in algorithm~\ref{alg:fwd-conv-layer};
% }

\For{$k'\in\{0,\cdots,\lceil\frac{\epsilon_l}{r}\rceil-1\}$}{
        \Comment{iterating each filter group}\;
        \For{$(u,~v)\in\{0,\cdots,\tilde{\gamma}_{l+1}-1\}^2$}{
                $\hat{C}^{(l+1)}_{k',u,v}\leftarrow 0$; 
                \Comment{initialize each output}\;
                $(u',~v')\leftarrow (\delta_l\cdot u,~\delta_l\cdot v)$\;
                \For{$(x,~y)\in\{0,\cdots,\gamma_l-1\}^2$}{
                        \For{$i\in\{0,\cdots,\alpha_l-1\}$}{ 
                            $\hat{C}^{(l+1)}_{k',u,v}\oplus = \hat{C}^{(l)}_{i,u'+x,v'+y}\otimes\hat{F}^{(l,k')}_{i,x,y}$;
                        }
                }
                $\hat{C}^{(l+1)}_{k',u,v}\leftarrow \hat{C}^{(l+1)}_{k',u,v}\otimes\hat{C}^{(l+1)}_{k',u,v}$; 
                \Comment{activation}
        }
}

\end{algorithm}

% \begin{algorithm}[htb]
% \caption{Propagation Through Convolutional Layer $l$ with Cross Filters Packing}
% \label{alg:fwd-conv-layer-cross-filters}

% \begin{algorithmic}[1]

% \For{$k\in\{0,\cdots,\lceil\frac{\epsilon_l}{r}\rceil-1\}$} \Comment{each cross filters packing}
%   \State Same as Lines $ 2-10$ in algorithm~\ref{alg:fwd-conv-layer}
% \EndFor

% \end{algorithmic}
% \end{algorithm}

% \input{integrity}

\section{Evaluation}
\label{sec:eval}

We implement the LHE-based inference scheme 
and run it on a moderate platform, i.e., a computer with Intel 2.6 GHz CPU that runs Ubuntu 20.04, 
for performance evaluation.

\subsection{CNN Models and LHE Parameters}
\label{subsec::cnn-strucutres}

We use the CNN models described in Section~\ref{subsec:cnn-model} 
with $c\in\{1,2,3,4\}$ and 
$f\in\{1,2\}$.
In the following, 
we use {\it CNN c-f} (e.g., {\it CNN 2-1}) to denote a CNN model 
with $c$ convolutional layers (CLs) and $f$ fully-connected layers (FLs). 
All experiments use the MNIST dataset~\cite{lecun1998mnist} to 
classify the ten handwritten digits and each input image of $28\times28$.

We employ 
CKKS~\cite{cheon2017homomorphic} of the SEAL library~\cite{sealcrypto} as the LHE scheme. 
Different polynomial modulus $N$ are used to match different CNN configurations. 
Each plaintext has ${\cal S}=\frac{N}{2}$ slots encoding up to ${\cal S}$ values. 
We adopt $N=8192$ and $N=16384$ which support 
a maximal coefficient modulus bit length of 218 and 438, respectively, to 
attain a 128-bit security level~\cite{laine2017simple}. 
Since each CL or FL (including square activation) requires 
two levels of homomorphic multiplication, 
we adjust the polynomial modulus and coefficient modulus to 
make the LHE scheme to support different settings of CNNs.
The detailed parameters 
are listed in Table~\ref{tab:HE-paras}.

\begin{table}[htb]
\caption{The SEAL parameters for different CNN structures}
\label{tab:HE-paras}
\resizebox{\linewidth}{!}{
$\displaystyle
\begin{tabular}{|l|l|l|l|}
\hline
CNN models   & N     & Coefficient Modulus                     & Levels \\ \hline
\textit{CNN 1-2} & 8192  & \{34,30,30,30,30,30,34\}                & 6      \\ \hline
\textit{CNN 2-1} & 8192  & \{34,30,30,30,30,30,34\}                & 6      \\ \hline
\textit{CNN 2-2} & 16384 & \{40,30,30,30,30,30,30,40\}             & 7      \\ \hline
\textit{CNN 3-1} & 16384 & \{40,30,30,30,30,30,30,30,40\}          & 8      \\ \hline
\textit{CNN 3-2} & 16384 & \{40,30,30,30,30,30,30,30,30,40\}       & 9      \\ \hline
\textit{CNN 4-1} & 16384 & \{40,30,30,30,30,30,30,30,30,30,40\}    & 10     \\ \hline
\textit{CNN 4-2} & 16384 & \{40,30,30,30,30,30,30,30,30,30,30,40\} & 11     \\ \hline
\end{tabular}
$}
\end{table}

\subsection{Evaluation Results}

\subsubsection{Computation Time of LHE Operations} 

We first measure the computation time of 
$\oplus$, $\otimes$, {\it Rot} and {\it CMul} 
with $N=16384$ and various encryption levels. 
The results (averaged over 5000 measurements) 
are presented in Table~\ref{tab:FHE-operations}. 
% The results are recorded as the average computation time running each operation for 5000 times.
As we can see, 
$\otimes$ has the highest cost,
followed by $Rot$ whose cost is around 80\% of $\otimes$,
followed by $CMul$ whose cost is about 15-20\% of $\otimes$,
and $\oplus$ has the lowest cost. 
With the level increases, the costs for $\otimes$, $Rot$ and $CMul$ all increase linearly. 

% when the level is 6, 
% the cost for $\otimes$ is around 5 times as that of {\it CMul}, and 
% the the cost for {\it Rot} is slightly smaller than that of $\otimes$.
% Also, as the level increases,
% the costs for every operation is significantly increased. 
% When the level is 11, 
% the cost for each operation is much greater than the case of level number 6.  

\begin{table}[htb]
\centering
\caption{Execution time ($\mu s$) of LHE operations ($N=16384$)}
\label{tab:FHE-operations}
\resizebox{0.6\linewidth}{!}{
\begin{tabular}{|l|l|l|l|l|}
\hline
%                  & Add      & Mul           & Rot           & PMul \\ \hline
% \textit{CNN 1-2} & 126      & 11980         & 9279          & 2368 \\ \hline
% \textit{CNN 4-2} & 466      & 66271         & 54557         & 9507     \\ \hline
Level & $\oplus$ & $\otimes$ & \textit{Rot} & \textit{CMul} \\ \hline
2      & 93       & 6434      & 4542         & 1645          \\ \hline
3      & 127      & 10106     & 7311         & 2467          \\ \hline
4      & 172      & 14466     & 10719        & 3273          \\ \hline
5      & 209      & 19757     & 14995        & 4137          \\ \hline
6      & 253      & 25931     & 20057        & 5018          \\ \hline
7      & 298      & 33139     & 25916        & 5935          \\ \hline
8      & 345      & 39953     & 31722        & 6741          \\ \hline
9      & 397      & 49835     & 40167        & 7942          \\ \hline
10     & 443      & 57791     & 47144        & 8731          \\ \hline
11     & 498      & 68374     & 56366        & 9895          \\ \hline
\end{tabular}
}
\end{table}

\subsubsection{Scalability with Varying $n$ (number of simultaneously-available inputs)} 

We experiment with CNN 3-2 model to evaluate 
the impact of $n\in\{16,32,64,128,256,512\}$.
The cross-channel or the cross-filter optimizations are applied whenever possible
and the packing parameter $r$ is computed according to the afore-described design.
% then the number of packed ratio in the cross-filter and cross-channel optimization $r\in\{32,16,8,4,2,1\}$ correspondingly since the slots number $s=8192$ and the encoded input matrix has side $\tilde{\beta}_0=4$. 
% For the CLs, we apply the cross-filters and cross-filters alternatively to fully utilize the slots.

\begin{table*}[htb]
\centering
\caption{Computation Time for {\it CNN 3-2} (unit: second)}
\label{tab:parallel-images-delay}
\resizebox{\textwidth}{!}{
\begin{tabular}{|l|l|l|l|l|l|l|l|l|l|l|l|l|l|}
\hline
($n$, $r$)                      & Inputs        & Filters       & Weights       & CL1           & Square        & CL2           & Square        & CL3           & Square        & FL1           & FL2           &\textbf{Total}       & \textbf{Amortized}\\ \hline
(\textbf{16},32)                         & 8.076         & 2.967         & 4.668         & 22.535        & 2.013         & 15.008        & 0.916         & 0.920         & 0.018         & 3.560         & 0.704         & \textbf{45.674}               & \textbf{2.855} \\ \hline
(\textbf{32},16)                         &8.108          & 2.974         & 4.683         & 21.829        &1.943          & 13.780        & 0.886         & 0.927         & 0.018         & 3.236         & 0.692         & \textbf{43.311}               & \textbf{1.353}\\ \hline
(\textbf{64},8)                          &8.464          &6.188          &7.297          &44.310         &4.022          &24.790         &0.912          &1.740          &0.032          &3.835          &0.713          & \textbf{80.354}               & \textbf{1.256}\\ \hline
(\textbf{128},4)                         &8.409          &12.347         &12.170         &88.434         &8.074          &47.035         &0.880          &3.261          &0.058          &5.110          &0.690          & \textbf{153.542}              & \textbf{1.2}\\ \hline
(\textbf{256},2)                         &8.769          &25.088         &22.355         &177.42         &16.026         &91.092         &0.911          &6.676          &0.118          &8.65           &0.707          & \textbf{301.598}              & \textbf{1.178}\\ \hline
(\textbf{512},1)                         &8.742          &48.978         &41.712         &344.842        &32.435         &179.136        &0.885          &12.971         &0.224          &15.145         &0.696          & \textbf{586.334}              & \textbf{1.145}\\ \hline
\end{tabular}
}
\end{table*}

\begin{table}[htb]
\centering
\caption{Numbers of LHE operations at every stages of {\it CNN 1-2}. Input size: 4096 for CryptoNets; 64 for E2DM and our scheme. 
Each single value in inputs and parameters encryption represent the number of encryption times and each value in Square row represents the number of ciphertext-ciphertext multiplication operations. \textbf{The tuples represent the number of LHE operations in the order of $\oplus$, $\otimes$, $Rot$ and $CMult$. 
The 2-tuples, 3-tuples and 4-tuples match the first 2, 3 and 4 operations respectively.}
}
\label{tab:cnn-1-2-times}
\resizebox{\linewidth}{!}{
\begin{tabular}{|l|l|l|l|}
\hline
                      & Our Scheme        & E2DM                      & CryptoNets    \\ \hline
LHE Level             & 6                 & 8                        & 6           \\ \hline
Enc. Inputs         & 49                & 49                        & 784           \\ \hline
Enc. Filter         & 196               & 196                       & 196           \\ \hline %\cline{2-5} 
Enc. Weight1        & 256               & 4                         & 16384         \\ \hline %\cline{2-5} 
Enc. Weight2        & 64                & 1                         & 640           \\ \hline
Preprocess Weight1     &                   &  (1024,0,608,768)         &          \\ \hline %\cline{2-5} 
Preprocess Weight2     &                   &  (138,0,44,148)          &          \\ \hline
CL1                    & (192,196,0)       & (192,196,0)               & (12288,12544) \\ \hline %\cline{2-5} 
Square                 & 4                 & 4                         & 256           \\ \hline %\cline{2-5} 
FL1                    & (576,256,384)     & (512,256,320,256)         & (16320,16384) \\ \hline %\cline{2-5} 
Square                 & 64                & 1                         & 64            \\ \hline %\cline{2-5} 
FL2                    & (63,64,0)         & (77,10,29,64)            & (630,640)     \\ \hline %\cline{2-5} 
\textbf{Total}         & (831,584,384)     & (781,467,349,320)         & (29238,29888)     \\ \hline %\cline{2-5} 
\textbf{Amortized}     & (13,9.1,6)     & (12.2,7.3,5.5,5)         & (7.1,7.3)     \\ \hline
\end{tabular}

% \begin{tabular}{|l|l|l|l|}
% \hline
%                       & Our Scheme        & E2DM                      & CryptoNets    \\ \hline
% LHE Level             & 6                 & 8                        & 6           \\ \hline
% Enc. Inputs         & 49                & 49                        & 784           \\ \hline
% Enc. Filter         & 196               & 196                       & 196           \\ \hline %\cline{2-5} 
% Enc. Weight1        & 256               & 4                         & 16384         \\ \hline %\cline{2-5} 
% Enc. Weight2        & 64                & 1                         & 640           \\ \hline
% Preprocess Weight1     &                   &  (0,608)         &          \\ \hline %\cline{2-5} 
% Preprocess Weight2     &                   &  (0,44)          &          \\ \hline
% CL1                    & (196,0)       & (196,0)               & (12544,0) \\ \hline %\cline{2-5} 
% Square                 & 4                 & 4                         & 256           \\ \hline %\cline{2-5} 
% FL1                    & (256,384)     & (256,320)         & (16384,0) \\ \hline %\cline{2-5} 
% Square                 & 64                & 1                         & 64            \\ \hline %\cline{2-5} 
% FL2                    & (64,0)         & (10,29)            & (640,0)     \\ \hline %\cline{2-5} 
% \textbf{Total}         & (584,384)     & (467,349)         & (29888,0)     \\ \hline %\cline{2-5} 
% \textbf{Amortized}     & (9.1,6)     & (7.3,5.5)         & (7.3,0)     \\ \hline
% \end{tabular}

}
\end{table}

Table~\ref{tab:parallel-images-delay} presents the computation time for 
encoding/encrypting initial input data and model parameters (in the columns labelled {\em Inputs}, {\em Filters} and {\em Weights}),
and propagating through the layers (in the columns labelled {\em CL1-3}, {\em FL1-2}, and {\em Square} for activation function).
The table also shows the total time for each forward propagation round (in the column labelled {\em Total})
and the averaged time for each input image (in the column labelled {\em Amortized}). 
As we can see,
the total time increases with $n$ while 
the amortized time (i.e., the average time for processing each input image) decreases. 
This demonstrates the following trade-off: 
the smaller is the scale of simultaneously-available input, 
the shorter is the delay for obtaining the inference result 
(indicated by column {\em Total}) and 
the lower is the system throughput (indicated by column {\em Amortized}) 
Meanwhile, 
% we can also find that, 
the difference in amortized time is not significant when 
compared to the difference in the input size.
Specifically, as $n$ changes from $16$ to $512$, 
the amortized time changes only from $2.855s$ to $1.145s$.
This indicates that, our proposed design 
% (including the cross-channel and cross-filter optimizations)
has good {\em scalability} in terms of that 
{\em its performance increases with the size of simultaneously-available inputs
and meanwhile the performance does not degrade significantly when the input size is small}.

\subsubsection{Comparison with CryptoNets and E2DM}

We compare the performance of our scheme
with CryptoNets and E2DM, which are also based on LHE and the most related to our scheme. 
As the LHE systems employed by the compared schemes are not CKKS, which is used in our scheme,
and the reported evaluation results~\cite{jiang2018secure} for them
are obtained at different platforms, 
we do not directly compare in terms of measured computation time.
Instead, we choose to count the numbers of LHE operations that each of the schemes need to conduct when they are given the same input.
Also note that, E2DM supports only one convolutional layer, 
so comparison is conducted based on the CNN 1-2 model. 
Both E2DM and our scheme takes 64 images as parallel inputs,
while CryptoNets takes 4096 images as parallel inputs (as its best performance is attained when all the 4096 slots are used).

Table~\ref{tab:cnn-1-2-times} shows the main comparison results among the three schemes. 
Comparing CrytoNets with the other schemes,
CrytoNets has the highest total cost due to the large input size;
in fact, even when the number of simultaneously-available inputs is smaller,
CrytoNets still needs such cost since it is not adaptive to input size. 
When the size of simultaneously-available inputs is 4096, it has the lowest amortized cost.
Specifically, it needs only $7.3$ $\otimes$ operations for each input images, by average,
while E2DM needs $7.3$ $\otimes$, $5.5$ $Rot$ and $5$ $CMult$ operations, and
our scheme needs $9.1$ $\otimes$ and $6$ $Rot$ operations.

\begin{table}[htb]
\centering
\caption{Number of LHE operations at different levels. }
\label{tab:cnn-1-2-levels}
\resizebox{.85\linewidth}{!}{
\begin{tabular}{|c|cccc|cccc|}
\hline
\multicolumn{1}{|l|}{} & \multicolumn{4}{c|}{Our Scheme}                                                                             & \multicolumn{4}{c|}{E2DM}                                                                                          \\ \hline
Level           & \multicolumn{1}{c|}{$\oplus$} & \multicolumn{1}{c|}{$\otimes$} & \multicolumn{1}{c|}{\textit{Rot}} & \textit{CMul} & \multicolumn{1}{c|}{$\oplus$} & \multicolumn{1}{c|}{$\otimes$} & \multicolumn{1}{c|}{\textit{Rot}} & \textit{CMul} \\ \hline
7                      & \multicolumn{1}{c|}{-}        & \multicolumn{1}{c|}{-}         & \multicolumn{1}{c|}{-}            & -             & \multicolumn{1}{c|}{192}      & \multicolumn{1}{c|}{196}       & \multicolumn{1}{c|}{-}            & -             \\ \hline
6                      & \multicolumn{1}{c|}{-}        & \multicolumn{1}{c|}{-}         & \multicolumn{1}{c|}{-}            & -             & \multicolumn{1}{c|}{-}        & \multicolumn{1}{c|}{4}         & \multicolumn{1}{c|}{-}            & -             \\ \hline
5                      & \multicolumn{1}{c|}{192}      & \multicolumn{1}{c|}{196}       & \multicolumn{1}{c|}{-}            & -             & \multicolumn{1}{c|}{256}      & \multicolumn{1}{c|}{-}         & \multicolumn{1}{c|}{320}          & 256           \\ \hline
4                      & \multicolumn{1}{c|}{-}        & \multicolumn{1}{c|}{4}         & \multicolumn{1}{c|}{-}            & -             & \multicolumn{1}{c|}{256}      & \multicolumn{1}{c|}{256}       & \multicolumn{1}{c|}{-}            & -             \\ \hline
3                      & \multicolumn{1}{c|}{576}      & \multicolumn{1}{c|}{256}       & \multicolumn{1}{c|}{384}          & -             & \multicolumn{1}{c|}{-}        & \multicolumn{1}{c|}{1}         & \multicolumn{1}{c|}{-}            & -             \\ \hline
2                      & \multicolumn{1}{c|}{-}        & \multicolumn{1}{c|}{64}        & \multicolumn{1}{c|}{-}            & -             & \multicolumn{1}{c|}{64}       & \multicolumn{1}{c|}{-}         & \multicolumn{1}{c|}{26}           & 64            \\ \hline
1                      & \multicolumn{1}{c|}{63}       & \multicolumn{1}{c|}{64}        & \multicolumn{1}{c|}{-}            & -             & \multicolumn{1}{c|}{13}       & \multicolumn{1}{c|}{10}        & \multicolumn{1}{c|}{3}            & -             \\ \hline
\end{tabular}
}
\end{table}

However, the above comparison does not indicate that E2DM is more efficient than our scheme,
because it only counts the number of LHE operations but 
does not consider the impact of encryption level. 
Table~\ref{tab:cnn-1-2-levels} shows the distribution of the LHE operations at different levels. 
As we can see, 
the LHE operations of E2DM occur at higher levels that our scheme.
Together with the level-specific LHE operation costs shown in Table~\ref{tab:FHE-operations}, 
E2DM should incur higher cost.

As we can also see from Table~\ref{tab:cnn-1-2-times}, 
both CryptoNets and our scheme demand the highest encryption level of 6, 
while the level needed by E2DM is 8;
note that, as shown by Table~\ref{tab:FHE-operations}, 
higher level implies most costly LHE operations. 
Regarding input encryption, 
E2DM and our scheme use the same packing method when 
the input size is 64 and thus have the same cost; 
CryptoNets has higher cost due to large input size.
Regarding model encryption and preprocessing, 
our scheme has higher cost in encryption than E2DM, 
but E2DM needs to preprocess weight matrices which is not needed by our scheme;
nevertheless, these are just one-time cost. 

Overall, the comparisons indicate that,
{\em our scheme is more efficient than E2DM mainly due to the smaller encryption level that we need}.
Also, {\em CrytoNets has the lowest amortized cost but this is attainable only when a large number of inputs are simultaneously available}.

\subsubsection{Applicability to Different CNN Models}

To show the applicability of our model to generic CNN models, 
we evaluate our scheme with the CNN models whose configurations and parameters listed in section~\ref{subsec::cnn-strucutres}.
% and~\ref{subsec::fhe-params}. 
Due to space limit,
we show only the results of two settings in Tables~\ref{tab:cnn-2-1} and \ref{tab:cnn-4-2}. 
% consider different depth of CNN architectures. 
% The CNN structures and corresponding parameters are listed in section~\ref{subsec::cnn-strucutres} and~\ref{subsec::fhe-params}.

\begin{table}[htb]
\centering
\caption{LHE operations and time for {\it CNN 2-1} (Our scheme: n=16, r=16).}
\label{tab:cnn-2-1}
\resizebox{\linewidth}{!}{

\begin{tabular}{|l|ll|l|}
\hline
                     & \multicolumn{2}{c|}{LHE Operations}                           & \multicolumn{1}{c|}{Time}             \\ \hline
                     & \multicolumn{1}{l|}{Our Scheme}                  & \multicolumn{1}{l|}{CryptoNets}   & Our Scheme  \\ \hline
Enc.  Inputs             & \multicolumn{1}{l|}{225}                      & \multicolumn{1}{l|}{784}          & 3.183   \\ \hline %\cline{2-5}
Enc.  Filters            & \multicolumn{1}{l|}{149}                      & \multicolumn{1}{l|}{2384}         & 2.12  \\ \hline %\cline{2-7}
Enc.  Weights            & \multicolumn{1}{l|}{40}                       & \multicolumn{1}{l|}{640}           & 0.561   \\ \hline %\cline{2-7}
 \textbf{Total}      & \multicolumn{1}{l|}{189}                      & \multicolumn{1}{l|}{3024}         & 2.681  \\ \hline %\cline{2-7}
  CL1                & \multicolumn{1}{l|}{(1200,1225)}              & \multicolumn{1}{l|}{(92928,94864)}& 15.882 \\ \hline %\cline{2-7}
  Square             & \multicolumn{1}{l|}{25}                       & \multicolumn{1}{l|}{1936}          & 0.251    \\ \hline %\cline{2-7}
  CL2                & \multicolumn{1}{l|}{(112,100,16)}             & \multicolumn{1}{l|}{(25536,25600)}  & 0.859  \\ \hline %\cline{2-7}
  Square             & \multicolumn{1}{l|}{4}                        & \multicolumn{1}{l|}{64}            & 0.021   \\ \hline %\cline{2-7}
  FL1                & \multicolumn{1}{l|}{(110,40,80)}              & \multicolumn{1}{l|}{(630,640)}   & 0.258   \\ \hline %\cline{2-7}
 \textbf{Total}      & \multicolumn{1}{l|}{(1422,1394,96)}           & \multicolumn{1}{l|}{(119094,123104)}     &17.271 \\ \hline %\cline{2-7}
 \textbf{Amortized}  & \multicolumn{1}{l|}{(88.9,\textbf{87.1,6})}            & \multicolumn{1}{l|}{(29.1,\textbf{30.1})}  &1.079        \\ \hline
\end{tabular}
}
\end{table}

\begin{table}[htb]
\centering
\caption{LHE operations and time for {\it CNN 4-2} (Our Scheme: n=128, r=16).}
\label{tab:cnn-4-2}
\resizebox{\linewidth}{!}{
\begin{tabular}{|l|ll|l|}
\hline
                & \multicolumn{2}{c|}{LHE operations}                     & \multicolumn{1}{c|}{Time} \\ \hline
                & \multicolumn{1}{l|}{Our Scheme}         & CryptoNets      & Our Scheme                   \\ \hline
Enc. Inputs         & \multicolumn{1}{l|}{441}              & 784             & 20.036                     \\ \hline
Filters        & \multicolumn{1}{l|}{133}              & 2128            & 6.150                      \\ \hline %\cline{2-5} 
Weights        & \multicolumn{1}{l|}{320}              & 1184            & 14.471                     \\ \hline %\cline{2-5} 
\textbf{Total} & \multicolumn{1}{l|}{453}              & 3312            & 20.621                     \\ \hline
CL1            & \multicolumn{1}{l|}{(1944,2025,0)}    & (55296,57600)   & 133.093                    \\ \hline %\cline{2-5} 
Square         & \multicolumn{1}{l|}{81}               & 2304            & 4.579                      \\ \hline %\cline{2-5} 
CL2            & \multicolumn{1}{l|}{(2352,1764,784)}  & (57200,57600)   & 112.451                    \\ \hline %\cline{2-5} 
Square         & \multicolumn{1}{l|}{196}              & 400             & 7.368                      \\ \hline %\cline{2-5} 
CL3            & \multicolumn{1}{l|}{(315,324,0)}      & (8960,9216)     & 11.85                      \\ \hline %\cline{2-5} 
Square         & \multicolumn{1}{l|}{9}                & 256             & 0.222                      \\ \hline %\cline{2-5} 
CL4            & \multicolumn{1}{l|}{(48,36,16)}       & (2288,2304)     & 1.128                      \\ \hline %\cline{2-5} 
Square         & \multicolumn{1}{l|}{4}                & 16              & 0.059                      \\ \hline %\cline{2-5} 
FL1            & \multicolumn{1}{l|}{(576,256,384)}    & (960,1024)      & 5.452                      \\ \hline %\cline{2-5} 
FL2            & \multicolumn{1}{l|}{(63,64,0)}        & (150,160)       & 0.822                      \\ \hline %\cline{2-5} 
\textbf{Total} & \multicolumn{1}{l|}{(5298,4759,1184)} & (124854,130880) & 277.024                    \\ \hline %\cline{2-5} 
\textbf{Amortized}                & \multicolumn{1}{l|}{(41.4,\textbf{37.2,9.3})}                 &(15.2,\textbf{16})                 & 2.164                           \\ \hline
\end{tabular}
}
\end{table}

As in Table~\ref{tab:cnn-2-1},
when our scheme is applied to CNN 2-1 with $16$ parallel inputs,
to compute inference from the inputs takes $17.271$ seconds, which is 
$1.079$ seconds for each input. 
With the setting, 
our scheme performs $87.1$ $\otimes$ and $6$ $Rot$ operations for each of its $16$ input.
CryptoNets performs $30.1$ $\otimes$ for each of its $4096$ inputs; however,
if it has only $16$ inputs available, it performs $7705.6$ $\otimes$ operations per input,
which is around $82$ times higher.
Similar trends can be observed from Table~\ref{tab:cnn-4-2}, except that
the amortized cost increases as the scale of the model increases, which is reasonable. 
Hence, {\em our scheme has good applicability and scalability with varying size of simultaneously-available inputs under various CNN models}.

\section{Related Works}

We briefly discuss the related works as follows. 

% The works related to our proposed framework and scheme fall into the following categories.

% There are several researches focus on the privacy-preserving deep neural network inference. 
% Generally, these research works are mainly based on the following techniques: homomorphic encryption, multi-party computation or a combination of these techniques.

\subsection{LHE-based Schemes}

Schemes based on LHE were proposed in~\cite{gilad2016cryptonets,jiang2018secure,bourse2018fast,chou2018faster,dathathri2019chet,xie2019bayhenn,xie2021privacy}.
Among them, CryptoNets~\cite{gilad2016cryptonets} is one of the first works 
applying the packing technique~\cite{simd} for inference based on a CNN model. 
Assuming the simultaneous-availability of a large number of inputs, it
packs one value from each input into a ciphertext and thus process the inputs in parallel
to attain a high level of amortized efficiency.
E2DM~\cite{jiang2018secure} packs a matrix into a ciphertext and 
proposes an efficient algorithm to multiply two encrypted matrices; it also 
proposes packing and efficient processing for CNN models with one convolutional layer, but 
does not handle more general CNN models. 
PROUD~\cite{xie2021privacy} applies the techniques and 
introduces parallel execution to further speed up the system.
We also apply LHE to our proposed framework, but focus to
design the inference scheme applicable for more generic CNN models and
scalable to the availability of parallel inputs. 

\subsection{MPC-based Schemes}

Several recent works~\cite{mohassel2017secureml,rouhani2018deepsecure,liu2017oblivious,juvekar2018gazelle,mishra2020delphi,chandran2019ezpc,riazi2019xonn,riazi2018chameleon,lu2017crypti, xie2019bayhenn,xie2021privacy, huang2022cheetah} propose two-party computations schemes 
that require interactions between the client and server while the computation is being performed,
which could cause high communication overheads and latency.
Among them, for example, MiniONN~\cite{liu2017oblivious} and GAZELLE~\cite{juvekar2018gazelle} 
split the inputs so that both client and server hold additive secret shares of input, 
and garbled circuits are employed for non-linear computation.
DELPHI~\cite{mishra2020delphi} extends GAZELLE and designs a hybrid scheme that 
generates neural network architecture configurations to balance the trade-offs 
between performance and accuracy. Three-party computation-based schemes are also 
proposed in~\cite{mohassel2018aby3, wagh2020falcon, patra2020blaze, koti2021swift}
% but no-collusion among servers has to be assumed.
and four-party computation-based schemes are explored in~\cite{chaudhari2019trident, byali2019flash, koti2021swift}.
However, these three-party and four-party computation-based schemes all introduced under the assumption of a honesty-majority.

\subsection{TEE-based Schemes}

Schemes~\cite{tramer2018slalom, zhang2021citadel, natarajan2021chex} were also 
proposed based on TEE. 
Among them,  
% Recently, the TEE-based approach~\cite{tramer2018slalom, zhang2021citadel, natarajan2021chex} has been proposed on secure model training or inference.
% Salmon~\cite{tramer2018slalom} leverages Intel SGX enclave~\cite{mckeen2013innovative} to guarantee the inputs privacy and verify the integrity of linear computation in DNN inference. 
% Salmon is composed of a pre-processing phase and online phase where both are executed by trusted enclave and the online phase is collaborated with un-trusted server using additive secret sharing. 
% Yet, the model confidentiality is not protected.
Citadel~\cite{zhang2021citadel} preserves data and model privacy 
in distributed training by partitioning code to two parts, 
i.e., data handling code executed by multiple training enclaves and 
model handling code executed by an aggregation enclave.    
CHEX-MIX~\cite{natarajan2021chex} combines HE with TEE to 
protect data and model confidentiality as well as 
verify the computation integrity, when the client and server are mutually distrustful. 
The inference process are completely executed inside enclave with 
inputs encrypted with HE and attested model parameters.
We also adopt TEE in our proposed framework. However 
different from the related works, we minimize the involvement of TEE 
only for essential works such as LHE key management and 
decryption/re-encryption of final results; we assign the most workload to 
the REE of server which has more resources than the TEE.

\section{Conclusions}
\label{sec:conclusion}

In this paper, we propose a new framework based on symbiotic integration of LHE and TEE,
which enables collaboration among mutually-untrusted three parties.
% while
% minimizing the involvement of resource-constrained TEE and allowing the full utilization of 
% the untrusted but resource-rich part of server.
We also propose a generic and efficient LHE-based inference scheme,
along with optimizations, as an important performance-determining component of the framework. 
We have implemented the proposed system on a moderate platform, and 
conducted extensive evaluations to show that,
our proposed system is applicable and scalable to various settings,
and it has better or comparable performance when 
compared with the state-of-the-art solutions
which are more restrictive in applicability and scalability.

\bibliographystyle{IEEEtran}
\scriptsize\bibliography{main.bbl}

\end{document}